# CLEAR: <u>C</u>ross-<u>L</u>ayer <u>E</u>xploration for <u>A</u>rchitecting <u>R</u>esilience
## Combining Hardware and Software Techniques to Tolerate Soft Errors in Processor Cores

Eric Cheng[1], Shahrzad Mirkhani[1], Lukasz G. Szafaryn[2], Chen-Yong Cher[3], Hyungmin Cho[1],
Kevin Skadron[2], Mircea R. Stan[2], Klas Lilja[4], Jacob A. Abraham[5], Pradip Bose[3], Subhasish Mitra[1]

[1]Stanford University, [2]University of Virginia, [3]IBM Research, [4]Robust Chip, Inc., [5]University of Texas at Austin

## Abstract

We present a first of its kind framework which overcomes a major challenge in the design of digital systems that are resilient to reliability failures: achieve desired resilience targets at minimal costs (energy, power, execution time, area) by combining resilience techniques across various layers of the system stack (circuit, logic, architecture, software, algorithm). This is also referred to as *cross-layer resilience*. In this paper, we focus on radiation-induced soft errors in processor cores. We address both single-event upsets (SEUs) and single-event multiple upsets (SEMUs) in terrestrial environments. Our framework automatically and systematically explores the large space of comprehensive resilience techniques and their combinations across various layers of the system stack (586 cross-layer combinations in this paper), derives cost-effective solutions that achieve resilience targets at minimal costs, and provides guidelines for the design of new resilience techniques. We demonstrate the practicality and effectiveness of our framework using two diverse designs: a simple, in-order processor core and a complex, out-of-order processor core. Our results demonstrate that a carefully optimized combination of circuit-level hardening, logic-level parity checking, and micro-architectural recovery provides a highly cost-effective soft error resilience solution for general-purpose processor cores. For example, a 50× improvement in silent data corruption rate is achieved at only 2.1% energy cost for an out-of-order core (6.1% for an in-order core) with no speed impact. However, selective circuit-level hardening alone, guided by a thorough analysis of the effects of soft errors on application benchmarks, provides a cost-effective soft error resilience solution as well (with ~1% additional energy cost for a 50× improvement in silent data corruption rate).

## CCS Concepts

• **General and reference** → **Reliability**; • **Hardware** → **Fault tolerance**; Transient errors and upsets; • **Computer systems organization** → **Reliability**

## Keywords

Cross-layer resilience; soft errors

## 1. Introduction

This paper addresses the *cross-layer resilience challenge* for designing robust digital systems: given a set of resilience techniques at various abstraction layers (circuit, logic, architecture, software, algorithm), how does one protect a given design from radiation-induced soft errors using (perhaps) a **combination** of these techniques, **across multiple abstraction layers**, such that overall soft error resilience targets are met at minimal costs (energy, power, execution time, area)? Specific soft error resilience targets addressed in this paper are: *Silent Data Corruption* (*SDC*), where an error causes the system to output an incorrect result without error indication; and, *Detected but Uncorrected Error* (*DUE*), where an error is detected (e.g., by a resilience technique or a system crash or hang) but is not recovered automatically without user intervention.

The need for *cross-layer resilience*, where multiple error resilience techniques from different layers of the system stack cooperate to achieve cost-effective error resilience, is articulated in several publications (e.g., [Borkar 05, Cappello 14, Carter 10, DeHon 10, Gupta 14, Henkel 14, Pedram 12]).

There are numerous publications on error resilience techniques, many of which span multiple abstraction layers. These publications mostly describe specific implementations. Examples include structural integrity checking [Lu 82] and its derivatives (mostly spanning architecture and software layers) or the combined use of circuit hardening, error detection (e.g., using logic parity checking and residue codes) and instruction-level retry [Ando 03, Meaney 05, Sinharoy 11] (spanning circuit, logic, and architecture layers). Cross-layer resilience implementations in commercial systems are often based on "designer experience" or "historical practice." There exists no comprehensive framework to systematically address the cross-layer resilience challenge. Creating such a framework is difficult. It must encompass the entire design flow end-to-end, from comprehensive and thorough analysis of various combinations of error resilience techniques all the way to layout-level implementations, such that one can (automatically) determine which resilience technique or combination of techniques (either at the same abstraction layer or across different abstraction layers) should be chosen. However, such a framework is essential in order to answer important cross-layer resilience questions such as:

1. Is cross-layer resilience the best approach for achieving a given resilience target at low cost?

2. Are all cross-layer solutions equally cost-effective? If not, which cross-layer solutions are the best?

3. How do cross-layer choices change depending on application-level energy, latency, and area constraints?

4. How can one create a cross-layer resilience solution that is cost-effective across a wide variety of application workloads?

5. Are there general guidelines for new error resilience techniques to be cost-effective?

We present CLEAR (<u>C</u>ross-<u>L</u>ayer <u>E</u>xploration for <u>A</u>rchitecting <u>R</u>esilience), a first of its kind framework, which addresses the cross-layer resilience challenge. In this paper, we focus on the use of CLEAR for radiation-induced soft errors[1] in terrestrial environments.

Although the soft error rate of an SRAM cell or a flip-flop stays roughly constant or even decreases over technology generations, the system-level soft error rate increases with increased integration [Mitra 14, Seifert 10, 12]. Moreover, soft error rates can increase when lower supply voltages are used to improve energy efficiency [Mahatme 13, Pawlowski 14]. We focus on *flip-flop soft errors* because design techniques to protect them are generally expensive. Coding techniques are routinely used for protecting on-chip memories. Combinational logic circuits are significantly less susceptible to soft errors and do not pose a concern [Gill 09, Seifert 12]. We address both single-event upsets (*SEUs*) and single-event multiple upsets (*SEMUs*) [Lee 10, Pawlowski 14]. While CLEAR can address soft errors in various digital components of a complex System-on-a-Chip (including uncore components [Cho 15] and hardware accelerators), a detailed analysis of soft errors in all these components is beyond the scope of this paper. Hence, we focus on soft errors in processor cores.

To demonstrate the effectiveness and practicality of CLEAR, we explore 586 cross-layer combinations using ten representative error detection/correction techniques and four hardware error recovery techniques[2]. These techniques span various layers of the system stack: circuit, logic, architecture, software, and algorithm (Fig. 1). Our extensive cross-layer exploration encompasses over 9 million flip-flop soft error injections into two diverse processor core architectures (Table 1): a simple in-order SPARC Leon3 core (*InO-core*) and a complex super-scalar out-of-order Alpha IVM core (*OoO-core*), across 18 benchmarks: SPECINT2000 [Henning 00] and DARPA PERFECT [DARPA]. Such

---
[1] Other error sources (voltage noise and circuit-aging) may be incorporated into CLEAR, but are not the focus of this paper.
[2] An earlier version ([Cheng 16]) included combinations that were valid, but covered by cheaper combinations (e.g., LEAP-DICE + flush recovery vs. LEAP-DICE only). These extra combinations were removed as they do not change any conclusions.

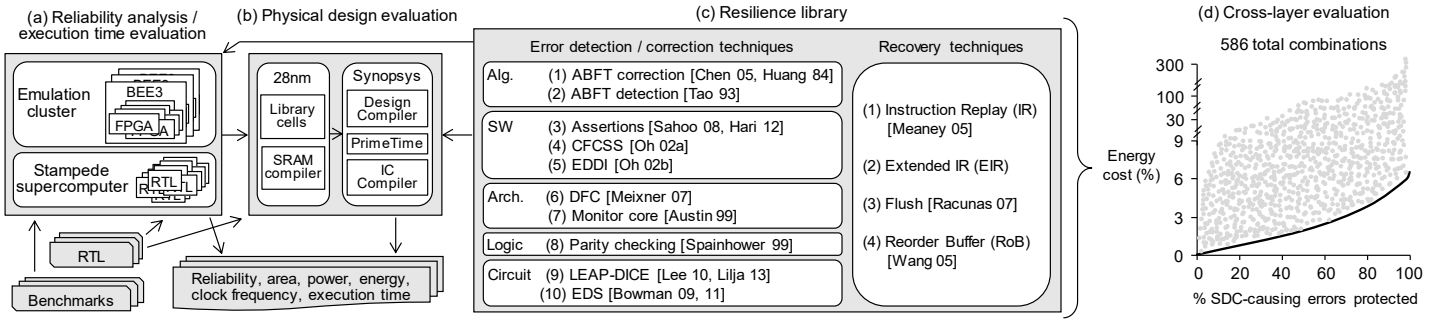

Figure 1. CLEAR Framework: (a) BEE3 emulation cluster / Stampede supercomputer injects over 9 million errors into two diverse processor architectures running 18 full-length application benchmarks. (b) Accurate physical design evaluation accounts for resilience overheads. (c) Comprehensive resilience library consisting of ten error detection / correction techniques + four hardware error recovery techniques. (d) Example illustrating thorough exploration of 586 cross-layer combinations with varying energy costs vs. percentage of SDC-causing errors protected.

extensive exploration enables us to conclusively answer the above cross-layer resilience questions:

1. For a wide range of error resilience targets, optimized cross-layer combinations can provide low cost solutions for soft errors.

2. Not all cross-layer solutions are cost-effective.

   a. For general-purpose processor cores, a carefully optimized combination of selective circuit-level hardening, logic-level parity checking, and micro-architectural recovery provides a highly effective cross-layer resilience solution. For example, a 50× SDC improvement (defined in Sec. 2.1) is achieved at 2.1% and 6.1% energy costs for the OoO- and InO-cores, respectively. The use of selective circuit-level hardening and logic-level parity checking is guided by a thorough analysis of the effects of soft errors on application benchmarks.

   b. When the application space can be restricted to matrix operations, a cross-layer combination of Algorithm Based Fault Tolerance (ABFT) correction, selective circuit-level hardening, logic-level parity checking, and micro-architectural recovery can be highly effective. For example, a 50× SDC improvement is achieved at 1.9% and 3.1% energy costs for the OoO- and InO-cores, respectively. But, this approach may not be practical for general-purpose processor cores targeting general applications.

   c. Selective circuit-level hardening, guided by a thorough analysis of the effects of soft errors on application benchmarks, provides a highly effective soft error resilience approach. For example, a 50× SDC improvement is achieved at 3.1% and 7.3% energy costs for the OoO- and InO-cores, respectively.

3. The above conclusions about cost-effective soft error resilience techniques largely hold across various application characteristics (e.g., latency constraints despite errors in soft real-time applications).

4. Selective circuit-level hardening (and logic-level parity checking) techniques are guided by the analysis of the effects of soft errors on application benchmarks. Hence, one must address the challenge of potential mismatch between application benchmarks vs. applications in the field, especially when targeting high degrees of resilience (e.g., 10× or more SDC improvement). We overcome this challenge using various flavors of circuit-level hardening techniques (details in Sec. 4).

5. Cost-effective resilience approaches discussed above provide bounds that new soft error resilience techniques must achieve to be competitive. It is, however, crucial that the benefits and costs of new techniques are evaluated thoroughly and correctly before publication.

Table 1. Processor designs studied.

| Core | Design | Description | Clk. freq. | Error injections | Instructions Per Cycle |
|---|---|---|---|---|---|
| InO | Leon3 [Leon] | Simple, in-order (1,250 flip-flops) | 2.0 GHz | 5.9 million | 0.4 |
| OoO | IVM [Wang 04] | Complex, super-scalar, out-of-order (13,819 flip-flops) | 600 MHz | 3.5 million | 1.3 |

---

[3] 11 SPEC / 7 PERFECT benchmarks for InO-cores and 8 SPEC / 3 PERFECT for OoO-cores (missing benchmarks contain floating-point instructions not executable by the OoO-core RTL model).

## 2. CLEAR Framework

Figure 1 gives an overview of the CLEAR framework. Individual components of the framework are discussed below.

### 2.1 Reliability Analysis

CLEAR is not merely an error rate projection tool; rather, reliability analysis is a component of the overall CLEAR framework.

We use flip-flop soft error injections for reliability analysis with respect to radiation-induced soft errors. This is because radiation test results confirm that injection of single bit-flips into flip-flops closely models soft error behaviors in actual systems [Bottoni 14, Sanda 08]. Furthermore, flip-flop-level error injection is crucial since naïve high-level error injections can be highly inaccurate [Cho 13]. For individual flip-flops, both SEUs and SEMUs manifest as single-bit errors. Our SEMU-tolerant circuit hardening and our layout implementations ensure this assumption holds for both the baseline and resilient designs.

We injected over 9 million flip-flop soft errors into the RTL of the processor designs using three BEE3 FPGA emulation systems and also using mixed-mode simulations on the Stampede supercomputer (TACC at The University of Texas at Austin) (similar to [Cho 13, Davis 09, Ramachandran 08, Wang 04]). This ensures that error injection results have less than a 0.1% margin of error with a 95% confidence interval per benchmark. Errors are injected uniformly into all flip-flops and application regions, to mimic real world scenarios.

The SPECINT2000 [Henning 00] and DARPA PERFECT [DARPA] benchmark suites are used for evaluation[3]. The PERFECT suite complements SPEC by adding applications targeting signal and image processing domains. We chose the SPEC workloads since the original publications corresponding to the resilience techniques used them for evaluation. We ran benchmarks in their entirety.

Flip-flop soft errors can result in the following outcomes [Cho 13, Michalak 12, Sanda 08, Wang 04, 07]: **Vanished -** normal termination and output files match error-free runs, **Output Mismatch (OMM) -** normal termination, but output files are different from error-free runs, **Unexpected Termination (UT) -** program terminates abnormally, **Hang -** no termination or output within 2× the nominal execution time, **Error Detection (ED) -** an employed resilience technique flags an error, but the error is not recovered using a hardware recovery mechanism.

Using the above outcomes, any error that results in OMM causes SDC (i.e., an *SDC-causing error*). Any error that results in UT, Hang, or ED causes DUE (i.e., a *DUE-causing error*). Note that, there are no ED outcomes if no error detection technique is employed. The resilience of a protected (new) design compared to an unprotected (original, baseline) design can be defined in terms of *SDC improvement* (Eq. 1a) or *DUE improvement* (Eq. 1b). The susceptibility of flip-flops to soft errors is assumed to be uniform across all flip-flops in the design (but this parameter is adjustable in our framework).

Resilience techniques that increase the execution time of an application or add additional hardware also increase the susceptibility of the design to soft-errors. To accurately account for this situation, we calculate, based on [Schirmeier 15], a correction factor γ (where γ ≥ 1), which is applied to ensure a fair and accurate comparison for all techniques[4]. Take for instance the monitor core technique; in our implementation, it increases the number of flip-flops in a resilient OoO-core by 38%. These extra flip-flops become additional locations for soft errors to occur. This results in a γ correction of 1.38 in order to account for the increased susceptibility of the design to soft errors. Techniques which increase execution time have a similar impact. For example, CFCSS incurs a 40.6% execution time impact; a corresponding γ correction of 1.41. A technique such as DFC, which increases flip-flop count (20%) and execution time (6.2%), would need a γ correction of 1.28 (1.2×1.062) since the impact is multiplicative (increased flip-flop count over an increased duration). The γ correction factor allows us to account for these increased susceptibilities for fair and accurate comparisons of all resilience techniques considered [Schirmeier 15]. SDC and DUE improvements with γ=1 can be back-calculated by multiplying the reported γ value in Table 3 and do not change our conclusions.

$$SDC\ improvement = \frac{(original\ OMM\ count)}{(new\ OMM\ count)} \times \gamma^{-1} \quad \text{(Eq. 1a)}$$

$$DUE\ improvement = \frac{(original\ (UT+Hang)\ count)}{(new\ (UT+Hang+ED)\ count)} \times \gamma^{-1} \quad \text{(Eq. 1b)}$$

Reporting SDC and DUE improvements allows our results to be agnostic to absolute error rates. Although we have described the use of error injection-driven reliability analysis, the modular nature of CLEAR allows us to swap in other approaches as appropriate (e.g., our error injection analysis could be substituted with techniques like [Mirkhani 15b], once they are properly validated).

As shown in Table 2, across our set of applications, not all flip-flops will have errors that result in SDC or DUE (errors in 19% of flip-flops in the InO-core and 39% of flip-flops in the OoO-core always vanish regardless of the application). The logic design structures (e.g., lowest hierarchical-level RTL component) these flip-flops belong to are listed in Appendix A. This phenomenon has been documented in the literature [Sullivan 16] and is due to the fact that errors that impact certain structures (e.g., branch predictor, trap status registers, etc.) have no effect on program execution or correctness. Additionally, this means that resilience techniques would not normally need to be applied to these flip-flops. However, for completeness, we also report design points which would achieve the maximum improvement possible, where resilience is added to every single flip-flop (including those with errors that always vanish). This maximum improvement point provides an upper bound for cost (given the possibility that for a future application, a flip-flop that currently has errors that always vanish may encounter an SDC- or DUE-causing error).

Table 2. Distribution of flip-flops with errors resulting in SDC and/or DUE over all benchmarks studied (specific flip-flop structures in Appendix A).

| Core | % FFs with SDC-causing errors | % FFs with DUE-causing errors | % FFs with both SDC- and DUE-causing errors |
|---|---|---|---|
| InO | 60.1% | 78.3% | 81.2% |
| OoO | 35.7% | 52.1% | 61% |

*Error Detection Latency* (the time elapsed from when an error occurs in the processor to when a resilience technique detects the error) is also an important aspect to consider. An end-to-end reliable system must not only detect errors, but also recover from these detected errors. Long detection latencies impact the amount of computation that needs to be recovered and can also limit the types of recovery that are capable of recovering the detected error (Sec. 2.4).

### 2.2 Execution Time

Execution time is estimated using FPGA emulation and RTL simulation. Applications are run to completion to accurately capture the execution time of an unprotected design. We also report the error-free execution time impact associated with resilience techniques at the architecture, software, and algorithm levels. For resilience techniques at the circuit and logic levels, our design methodology maintains the same clock speed as the unprotected design.

### 2.3 Physical Design

We used Synopsys design tools (Design Compiler, IC compiler, and Primetime) [Synopsys] with a commercial 28nm technology library (with corresponding SRAM compiler) to perform synthesis, place-and-route, and power analysis. *Synthesis and place-and-route* (*SP&R*) was run for all configurations of the design (before and after adding resilience techniques) to ensure all constraints of the original design (e.g., timing and physical design) were met for the resilient designs. Design tools often introduce artifacts (e.g., slight variations in the final design over multiple SP&R runs) that impact the final design characteristics (e.g., area, power). These artifacts can be caused by small variations in the RTL or optimization heuristics, for example. To account for these artifacts, we generated separate resilient designs based on error injection results for each individual application benchmark. SP&R was then performed for each of these designs, and the reported design characteristics were averaged to minimize the artifacts. For example, for each of our 18 application benchmarks, a separate resilient design that achieves a 50× SDC improvement using LEAP-DICE only is created. The costs to achieve this improvement are reported by averaging across the 18 designs. Relative standard deviation (i.e., standard deviation / mean) across all experiments range from 0.6-3.1%. Finally, we note that all layouts created during physically design are carefully generated in order to mitigate the impact of SEMUs (as explained in Sec. 2.4).

### 2.4 Resilience Library

We carefully chose ten error detection and correction techniques together with four hardware error recovery techniques. These techniques largely cover the space of existing soft error resilience techniques. The characteristics (e.g., costs, resilience improvement, etc.) of each resilience technique when used as a *standalone solution* (e.g., an error detection / correction technique by itself or, optionally, in conjunction with a recovery technique) are presented in Table 3.

**Circuit**: The *hardened flip-flops* (LEAP-DICE, Light Hardened LEAP, LEAP-ctrl) in Table 4 are designed to tolerate both SEUs and SEMUs at both nominal and near-threshold operating voltages [Lee 10, Lilja 13]. SEMUs especially impact circuit techniques since a single strike affects multiple nodes within a flip-flop. Thus, these specially designed hardened flip-flops, which tolerate SEMUs through charge cancellation, are required. Hardened flip-flops have been experimentally validated using radiation experiments on test chips fabricated in 90nm, 45nm, 40nm, 32nm, 28nm, 20nm, and 14nm nodes in both bulk and SOI technologies and can be incorporated into standard cell libraries (i.e., standard cell design guidelines are satisfied) [Lee 10, Lilja 13, Lilja 16, Quinn 15a, Quinn 15b, Turowski 15]. The LEAP-ctrl flip-flop is a special design, which can operate in resilient (high resilience, high power) and economy (low resilience, low power) modes. It is useful in situations where a software or algorithm technique only provides protection when running specific applications and thus, selectively enabling low-level hardware resilience when the former techniques are unavailable may be beneficial. While *Error Detection Sequential* (*EDS*) [Bowman 09, 11] was originally designed to detect timing errors, it can be used to detect flip-flop soft errors as well. While EDS incurs less overhead at the individual flip-flop level vs. LEAP-DICE, for example, EDS requires delay buffers to ensure minimum hold constraints, aggregation and routing of error detection signals to an output (or recovery module), and a recovery mechanism to correct detected errors. These factors can significantly increase the overall costs for implementing a resilient design utilizing EDS (Table 17).

---

[4] Research literature commonly considers γ=1. We report results using true γ values, but our conclusions hold for γ=1 as well (latter is optimistic).

Table 3. Individual resilience techniques: costs and improvements when implemented as a standalone solution.

| Layer | Technique | | Area cost | Power cost | Energy cost | Exec. time impact | Avg. SDC improve | Avg. DUE improve | False positive | Detection latency | γ |
|---|---|---|---|---|---|---|---|---|---|---|---|
| Circuit[5] | LEAP-DICE (no additional recovery needed) | InO | 0-9.3% | 0-22.4% | 0-22.4% | 0% | 1× - 5,000×[6] | 1× - 5,000×[6] | 0% | n/a | 0 |
| | | OoO | 0-6.5% | 0-9.4% | 0-9.4% | | | | | | |
| | EDS (without recovery - unconstrained) | InO | 0-10.7% | 0-22.9% | 0-22.9% | 0% | 1× - 100,000×[6] | 0.1×[6] - 1× | 0% | 1 cycle | 0 |
| | | OoO | 0-12.2% | 0-11.5% | 0-11.5% | | | | | | |
| | EDS (with IR recovery) | InO | 0-16.7% | 0-43.9% | 0-43.9% | 0% | 1× - 100,000×[6] | 1× - 100,000×[6] | 0% | 1 cycle | 1.4 |
| | | OoO | 0-12.3% | 0-11.6% | 0-11.6% | | | | | | 1.06 |
| Logic[5] | Parity (without recovery - unconstrained) | InO | 0-10.9% | 0-23.1% | 0-23.1% | 0% | 1× - 100,000×[6] | 0.1×[6] - 1× | 0% | 1 cycle | 0 |
| | | OoO | 0-14.1% | 0-13.6% | 0-13.6% | | | | | | |
| | Parity (with IR recovery) | InO | 0-26.9% | 0-44% | 0-44% | 0% | 1× - 100,000×[6] | 1× - 100,000×[6] | 0% | 1 cycle | 1.4 |
| | | OoO | 0-14.2% | 0-13.7% | 0-13.7% | | | | | | 1.06 |
| Arch. | DFC (without recovery - unconstrained) | InO | 3% | 1% | 7.3% | 6.2% | 1.2× | 0.5× | 0% | 15 cycles | 1.28 |
| | | OoO | 0.2% | 0.1% | 7.2% | 7.1% | | | | | 1.09 |
| | DFC (with EIR recovery) | InO | 37% | 33% | 41.2% | 6.2% | 1.2× | 1.4× | 0% | 15 cycles | 1.48 |
| | | OoO | 0.4% | 0.2% | 7.3% | 7.1% | | | | | 1.14 |
| | Monitor core (with RoB recovery) | OoO | 9% | 16.3% | 16.3% | 0% | 19× | 15× | 0% | 128 cycles | 1.38 |
| Software[7] | Software assertions for general-purpose processors (without recovery - unconstrained) | InO | 0% | 0% | 15.6% | 15.6%[8] | 1.5×[9] | 0.6× | 0.003% | 9.3M cycles[10] | 1.16 |
| | CFCSS (without recovery - unconstrained) | InO | 0% | 0% | 40.6% | 40.6% | 1.5× | 0.5× | 0% | 6.2M cycles[10] | 1.41 |
| | EDDI (without recovery - unconstrained) | InO | 0% | 0% | 110% | 110% | 37.8×[11] | 0.3× | 0% | 287K cycles[10] | 2.1 |
| Alg. | ABFT correction (no additional recovery needed) | InO OoO | 0% | 0% | 1.4% | 1.4% | 4.3× | 1.2× | 0% | n/a | 1.01 |
| | ABFT detection (without recovery - unconstrained) | InO OoO | 0% | 0% | 24% | 1-56.9%[12] | 3.5× | 0.5× | 0% | 9.6M cycles[10] | 1.24 |

Table 4. Resilient flip-flops.

| Type | Soft Error Rate | Area | Power | Delay | Energy |
|---|---|---|---|---|---|
| Baseline | 1 | 1 | 1 | 1 | 1 |
| Light Hardened LEAP (LHL) | $2.5 \times 10^{-1}$ | 1.2 | 1.1 | 1.2 | 1.3 |
| LEAP-DICE | $2.0 \times 10^{-4}$ | 2.0 | 1.8 | 1 | 1.8 |
| LEAP-ctrl (economy mode) | 1 | 3.1 | 1.2 | 1 | 1.2 |
| LEAP-ctrl (resilient mode) | $2.0 \times 10^{-4}$ | 3.1 | 2.2 | 1 | 2.2 |
| EDS[13] | ~100% detect | 1.5 | 1.4 | 1 | 1.4 |

**Logic**: *Parity checking* provides error detection by checking flip-flop inputs and outputs [Spainhower 99]. Our design heuristics reduce the cost of parity while also ensuring that clock frequency is maintained as in the original design (by varying the number of flip-flops checked together, grouping flip-flops by timing slack, pipelining parity checker logic, etc.) Naïve implementations of parity checking can otherwise degrade design frequency by up to 200 MHz (20%) or increase energy cost by 80% on the InO-core. We minimize SEMUs through layouts that ensure a minimum spacing (the size of one flip-flop) between flip-flops checked by the same parity checker. This ensures that only one flip-flop, in a group of flip-flops checked by the same parity checker, will encounter an upset due to a single strike in our 28nm technology in terrestrial environments [Amusan 09]. Although a single strike could impact multiple flip-flops, since these flip-flops are checked by different checkers, the upsets will be detected. Since this absolute minimum spacing will remain constant, the relative spacing required between flip-flops will increase at smaller technology nodes, which may exacerbate the difficulty of implementation. Minimum spacing is enforced by applying design constraints during the layout stage. This constraint is important because even in large designs, flip-flops will still tend to be placed very close to one another. Table 5 shows the distribution of distances that each flip-flop has to its next nearest neighbor in a baseline design (this does not correspond to the spacing between flip-flops checked by the same logic parity checker). As shown, the majority of flip-flops are actually placed such that they would be susceptible to a SEMU. After applying parity checking, we see that no flip-flop, within a group checked by the same parity checker, is placed such that it will be vulnerable to a SEMU (Table 6).

Table 5. Distribution of spacing between a flip-flop and its nearest neighbor in a baseline (original, unprotected) design.

| Distance | InO-core | OoO-core |
|---|---|---|
| < 1 flip-flop length away (i.e., flip-flops are adjacent and vulnerable to a SEMU) | 65.2% | 42.2% |
| 1 - 2 flip-flop lengths away | 30% | 30.6% |
| 2 - 3 flip-flop lengths away | 3.7% | 18.4% |
| 3 - 4 flip-flop lengths away | 0.6% | 3.5% |
| > 4 flip-flop lengths away | 0.5% | 5.3% |

Table 6. Distribution of spacing between a flip-flop and its nearest neighbor in the same parity group (i.e., minimum distance between flip-flops checked by the same parity checker).

| Distance | InO-core | OoO-core |
|---|---|---|
| < 1 flip-flop length away (i.e., flip-flops are adjacent and vulnerable to a SEMU) | 0% | 0% |
| 1 - 2 flip-flop lengths away | 7.8% | 8.8% |
| 2 - 3 flip-flop lengths away | 5.3% | 10.6% |
| 3 - 4 flip-flop lengths away | 3.4% | 18.3% |
| > 4 flip-flop lengths away | 83.3% | 62.2% |
| Average distance | 4.4 flip-flops | 12.8 flip-flops |

---

[5] Circuit and logic techniques have tunable costs/resilience (e.g., for InO-cores, 5× SDC improvement using LEAP-DICE is achieved at 4.3% energy cost while 50× SDC improvement is achieved at 7.3% energy cost). This is achievable through selective insertion guided by error injection using application benchmarks.

[6] Maximum improvement reported is achieved by protecting every single flip-flop in the design.

[7] Software techniques are generated for InO-cores only since the LLVM compiler no longer supports the Alpha architecture.

[8] Some software assertions for general-purpose processors (e.g., [Sahoo 08]) suffer from false positives (i.e., an error is reported during an error-free run). The execution time impact reported discounts the impact of false positives.

[9] Improvements differ from previous publications that injected errors into architectural registers. [Cho 13] demonstrated such injections can be highly inaccurate; we used highly accurate flip-flop-level error injections.

[10] Actual detection latency for software and algorithm techniques may be shorter in practice. On our emulation platforms, measured detection latencies includes the time to trap and cleanly exit execution (on the order of a few thousand cycles).

[11] We report results for EDDI with store-readback [Lin 14]. Without this enhancement, EDDI provides 3.3× SDC / 0.4× DUE improvement.

[12] Execution time impact for ABFT detection can be high since error detection checks may require computationally-expensive calculations.

[13] For EDS, the costs are listed for the flip-flop only. Error signal routing and delay buffers (included in Table 3) increase overall cost [Bowman 11].

Logic parity is implemented using an XOR-tree based predictor and checker, which detects flip-flops soft errors. This implementation differs from logic parity prediction, which also targets errors inside combinational logic [Mitra 00]. XOR-tree logic parity is sufficient for detecting flip-flop soft errors (with the minimum spacing constraint applied). "Pipelining" in the predictor tree (Fig. 2) may be required to ensure 0% clock period impact. We evaluated the following heuristics for forming *parity groups* (the specific flip-flops that are checked together) to minimize cost of parity (cost comparisons in Table 7):

1. Parity group size: flip-flops are clustered into a constant power of 2-sized group, which amortizes the parity logic cost by allowing the use of full binary trees at the predictor and checker. The last set of flip-flops will consist of modulo "group size" of flip-flops.

2. Vulnerability: flip-flops are sorted by decreasing susceptibility to errors causing SDC or DUE and grouped into a constant power of 2-sized group. The last set of flip-flops will consist of modulo "group size" of flip-flops.

3. Locality: flip-flops are grouped by their location in the layout, in which flip-flops in the same functional unit are grouped together to help reduce wire routing for the predictor and checker logic. A constant power of 2-sized groups are formed with the last group consisting of modulo "group size" of flip-flops.

4. Timing: flip-flops are sorted based on their available timing path slack and grouped into a constant power of 2-sized group. The last set of flip-flops will consist of modulo "group size" of flip-flops.

5. Optimized: Fig. 3 describes our heuristic. Our solution is the most optimized and is the configuration we use to report overhead values.

When unpipelined parity can be used, it is better to use larger-sized groups (e.g., 32-bit groups) in order to amortize the additional predictor/checker logic to the number of flip-flops protected. However, when pipelined parity is required, we found 16-bit groups to be a good option. This is because beyond 16-bits, additional pipeline flip-flops begin to dominate costs. These factors have driven our implementation of the previously described heuristics.

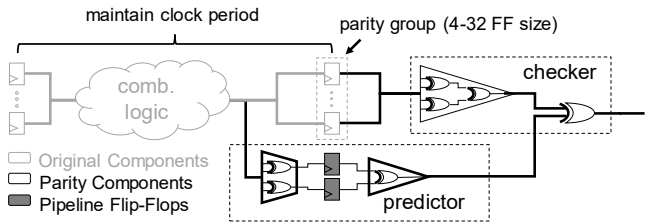

Figure 2. "Pipelined" logic parity.

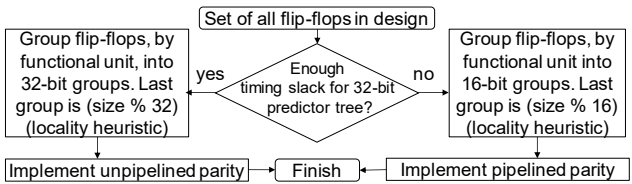

Figure 3. Logic parity heuristic for low cost parity implementation. 32-bit unpipelined parity and 16-bit pipelined parity were experimentally determined to be the lowest cost configurations.

Table 7. Comparison of heuristics for "pipelined" logic parity implementations to protect all flip-flops on the InO-core.

| Heuristic | Area cost | Power cost | Energy cost |
|---|---|---|---|
| Vulnerability(4-bit parity group) | 15.2% | 42% | 42% |
| Vulnerability(8-bit parity group) | 13.4% | 29.8% | 29.8% |
| Vulnerability(16-bit parity group) | 13.3% | 27.9% | 27.9% |
| Vulnerability(32-bit parity group) | 14.6% | 35.3% | 35.3% |
| Locality (16-bit parity group) | 13.4% | 29.4% | 29.4% |
| Timing (16-bit parity group) | 11.5% | 26.8% | 26.8% |
| Optimized (16-/32-bit groups) | 10.9% | 23.1% | 23.1% |

**Architecture**: Our implementation of *Data Flow Checking* (*DFC*), which checks static dataflow graphs, includes *Control Flow Checking (CFC)*, which checks static control-flow graphs. This combination checker resembles that of [Meixner 07], which is also similar to the checker in [Lu 82].

Compiler optimization allows us to embed the static signatures required by the checkers into unused delay slots in the software, thereby reducing execution time overhead by 13%.

Table 8 helps explain why DFC is unable to provide high SDC and DUE improvement. Of flip-flops that have errors that result in SDCs and DUEs (Sec. 2.1), DFC checkers detect SDCs and DUEs in less than 68% of these flip-flops (these 68% of flip-flops are distributed across all pipeline stages). For these 68% of flip-flops, on average, DFC detects less than 40% of the errors that result in SDCs or DUEs. This is because not all errors that result in an SDC or DUE will corrupt the dataflow or control flow signatures checked by the technique (e.g., register contents are corrupted and written out to a file, but the executed instructions remain unchanged). The combination of these factors means DFC is only detecting ~30% of SDCs or DUEs; thus, the technique provides low resilience improvement. These results are consistent with previously published data (detection of ~16% of non-vanished errors) on the effectiveness of DFC checkers in simple cores [Meixner 07].

Table 8. DFC error coverage.

| | InO | | OoO | |
|---|---|---|---|---|
| | SDC | DUE | SDC | DUE |
| % flip-flops with a SDC- / DUE-causing error that are detected by DFC | 57% | 68% | 65% | 66% |
| % of SDC- / DUE-causing errors detected (average per FF that is protected by DFC) | 30% | 30% | 29% | 40% |
| Overall % of SDC- / DUE-causing errors detected (for all flip-flops in the design) | 15.9% | 27% | 19.3% | 30% |
| Resulting improvement (Eq. 1) | 1.2× | 1.4× | 1.2× | 1.4× |

*Monitor cores* are checker cores that validate instructions executed by the main core (e.g., [Austin 99, Lu 82]). We analyze monitor cores similar to [Austin 99]. For InO-cores, the size of the monitor core is of the same order as the main core, and hence, excluded from our study. For OoO-cores, the simpler monitor core can have lower throughput compared to the main core and thus stall the main core. We confirm (via IPC estimation) that our monitor core implementation does not stall the main core (Table 9).

Table 9. Monitor core vs. main core.

| Design | Clk. freq. | Average Instructions Per Cycle (IPC) |
|---|---|---|
| OoO-core | 600 MHz | 1.3 |
| Monitor core | 2 GHz | 0.7 |

**Software**: *Software assertions for general-purpose processors* check program variables to detect errors. We combine assertions from [Hari 12, Sahoo 08] to check both data and control variables to maximize error coverage. Checks for data variables (e.g., end result) are added via compiler transformations using training inputs to determine the valid range of values for these variables (e.g., likely program invariants). Since such assertion checks are added based on training inputs, it is possible to encounter false positives, where an error is reported in an error-free run. We have determined this false positive rate by training the assertions using representative inputs. However, we perform final analysis by incorporating the input data used during evaluation into the training step in order to give the technique the best possible benefit and to eliminate the occurrence of false positives. Checks for control variables (e.g., loop index, stack pointer, array address) are determined using application profiling and are manually added in the assembly code.

In Table 10, we breakdown the contribution to cost, improvement, and false positives resulting from assertions checking data variables [Sahoo 08] vs. those checking control variables [Hari 12]. Table 11 demonstrates the importance of evaluating resilience techniques using accurate injection ([Cho 13])[14]. Depending on the particular error injection model used, SDC improvement could be over-estimated for one benchmark and under-estimated for another. For instance, using inaccurate architecture register error injection (regU), one would be led to believe that software assertions provide 3× the SDC improvement than they do in reality (e.g., when evaluated using flip-flop-level error injection).

In order to pinpoint the sources of inaccuracy between the actual improvement rates that were determined using accurate flip-flop-level error injection vs. those published in the literature, we conducted error injection campaigns at other levels of abstraction (architecture register and program variable). However, even then, we were unable to exactly reproduce previously published improvement rates. Some additional differences in our architecture and program variable injection methodology compared to the published methodology may account for this discrepancy:

1. Our architecture register and program variable evaluations were conducted on a SPARCv8 in-order design rather than a SPARCv9 out-of-order design.

2. Our architecture register and program variable methodology injects errors uniformly into all program instructions while previous publications chose to only inject into integer instructions of floating-point benchmarks.

3. Our architecture register and program variable methodology injects errors uniformly over the full application rather than injecting only into the core of the application during computation.

4. Since our architecture register and program variable methodology injects errors uniformly into all possible error candidates (e.g., all cycles and targets), the calculated improvement covers the entire design. Previous publications calculated improvement over the limited subset of error candidates (out of all possible error candidates) that were injected into and thus, only covers a subset of the design.

Table 10. Comparison of assertions checking data (e.g., end result) vs. control (e.g., loop index) variables.

|  | Data variable check | Control variable check | Combined check |
|---|---|---|---|
| Execution time impact | 12.1% | 3.5% | 15.6% |
| SDC improvement | 1.5× | 1.1× | 1.5× |
| DUE improvement | 0.7× | 0.9× | 0.6× |
| False positive rate | 0.003% | 0% | 0.003% |

Table 11. Comparison of SDC improvement and detection for assertions when injecting errors at various levels.

| App.[14] | Flip-flop (ground truth) | Register uniform (regU) | Register write (regW) | Program variable uniform (varU) | Program variable write (varW) |
|---|---|---|---|---|---|
| bzip2 | 1.8× | 1.6× | 1.1× | 1.9× | 1.5× |
| crafty | 0.5× | 0.3× | 0.5× | 0.7× | 1.1× |
| gzip | 2× | 19.3× | 1× | 1.6× | 1.1× |
| mcf | 1.1× | 1.3× | 0.9× | 1× | 1.8× |
| parser | 2.4× | 1.7× | 1× | 2.4× | 2× |
| avg. | 1.6× | 4.8× | 0.9× | 1.5× | 1.5× |

*Control Flow Checking by Software Signatures* (*CFCSS*) checks static control flow graphs and is implemented via compiler modification similar to [Oh 02a]. We can analyze CFCSS in further detail to gain deeper understanding as to why improvement for the technique is relatively low (Table 12). Compared to DFC (a technique with a similar concept), we see that CFCSS offers slightly better SDC improvement. However, since CFCSS only checks control flow signatures, many SDCs will still escape (e.g., the result of an add is corrupted and written to file). Additionally, certain DUEs, such as those which may cause a program crash, will not be detectable by CFCSS, or other software techniques, since execution may abort before a corresponding software check can be triggered. The relatively low resilience improvement using CFCSS has been corroborated in actual systems as well [Lovellette 02].

---

[14] We studied the same SPEC applications evaluated in [Sahoo 08].

Table 12. CFCSS error coverage.

|  | SDC | DUE |
|---|---|---|
| % flip-flops with a SDC- / DUE-causing error that is detected by CFCSS | 55% | 66% |
| % of SDC- / DUE-causing errors that are detected per FF that is protected by CFCSS | 61% | 14% |
| Resulting improvement (Eq. 1) | 1.5× | 0.5× |

*Error Detection by Duplicated Instructions* (*EDDI*) provides instruction redundant execution via compiler modification [Oh 2b]. We utilize EDDI with store-readback [Lin 14] to maximize coverage by ensuring that values are written correctly. From Table 13, it is clear why store-readback is important for EDDI. In order to achieve high SDC improvements, nearly all SDC causing errors need to be detected. By detecting an additional 12% of SDCs, store-readback increases SDC improvement of EDDI by an order of magnitude. Virtually all escaped SDCs are caught by ensuring that the values being written to the output are indeed correct (by reading back the written value). However, given that some SDC- or DUE-causing errors are still not detected by the technique, the results show that using naïve high-level injections will still yield incorrect conclusions (Table 14). Enhancements to EDDI such as Error detectors [Pattabiraman 09] and reliability-aware transforms [Rehman 14], are intended to reduce the number of EDDI checks (i.e., selective insertion of checks) in order to minimize execution time impact while maintaining high overall error coverage. We evaluated the Error detectors technique using flip-flop-level error injection and found that they provide an SDC improvement of 2.6× improvement (a 21% reduction in SDC improvement as compared to EDDI without store-readback). However, Error detectors requires software path tracking to recalculate important variables, which introduced a 3.9× execution time impact, greater than that of the original EDDI technique. The overhead corresponding to software path tracking can be reduced by implementing path tracking in hardware (as was done in the original work), but doing so eliminates the benefits of EDDI as a software-only technique.

Table 13. EDDI: importance of store-readback.

|  | SDC improvement | % SDC errors detected | SDC errors escaped | DUE improvement | % DUE errors detected | DUE errors escaped |
|---|---|---|---|---|---|---|
| Without store-readback | 3.3× | 86.1% | 49 | 0.4× | 19% | 3090 |
| With store-readback | 37.8× | 98.7% | 6 | 0.3× | 19.8% | 3006 |

Table 14. Comparison of SDC improvement and detection for EDDI when injecting errors at various levels (without store-readback).

| Injection location | SDC improvement | % SDC detected |
|---|---|---|
| Flip-flop (ground truth) | 3.3× | 86.1% |
| Register Uniform (RegU) | 2.0× | 48.8% |
| Register Write (RegW) | 6.6× | 84.8% |
| Program Variable Uniform (VarU) | 12.6× | 92.1% |
| Program Variable Write (VarU) | 100,000× | 100% |

**Algorithm**: *Algorithm Based Fault Tolerance (ABFT)* can detect (*ABFT detection*) or detect and correct errors (*ABFT correction*) through algorithm modifications [Bosilca 09, Chen 05, Huang 84, Nair 90]. Although ABFT correction algorithms can be used for detection-only (with minimally reduced execution time impact), ABFT detection algorithms cannot be used for correction. There is often a large difference in execution time impact between ABFT algorithms as well depending on the complexity of check calculation required. An ABFT correction technique for matrix inner product, for example, requires simple modular checksums (e.g., generated by adding all elements in a matrix row) – an inexpensive computation. On the other hand, ABFT detection for FFT, for example, requires expensive calculations using Parseval's theorem [Reddy 90]. For the particular applications we studied, the algorithms that were protected using ABFT detection often required more computationally-expensive checks than algorithms that were protected using ABFT correction; therefore, the former generally had greater execution time impact (relative to each of their own original baseline execution times). An additional complication arises when an ABFT detection-only algorithm is implemented. Due to the long error detection latencies imposed by ABFT detection (9.6 million cycles, on average), hardware recovery techniques are not feasible and higher level recovery mechanisms will impose significant overheads.

**Recovery**: We consider two recovery scenarios: *bounded latency*, i.e., an error must be recovered within a fixed period of time after its occurrence, and *unconstrained*, i.e., where no latency constraints exist and errors are recovered externally once detected (no hardware recovery is required). Bounded latency recovery is achieved using one of the following hardware recovery techniques (Table 15): *flush* or *reorder buffer* (*RoB*) recovery (both of which rely on flushing non-committed instructions followed by re-execution) [Racunas 07, Wang 05]; *instruction replay* (*IR*) or *extended instruction replay* (*EIR*) recovery (both of which rely on instruction checkpointing to rollback and replay instructions) [Meaney 05]. EIR is an extension of IR with additional buffers required by DFC for recovery. Flush and RoB are unable to recover from errors detected after the memory write stage of InO-cores or after the reorder buffer of OoO-cores, respectively (these errors will have propagated to architecture visible states). Hence, LEAP-DICE is used to protect flip-flops in these pipeline stages when using flush/RoB recovery. IR and EIR can recover detected errors in any pipeline flip-flop. IR recovery is shown in in Fig. 4 and flush recovery is shown in Fig. 5. Since recovery hardware serves as single points of failure, flip-flops in the recovery hardware itself needs to be capable of error correction (e.g., protected using hardened flip-flops when considering soft errors).

Table 15. Hardware error recovery costs.

| Core | Type | Area | Power | Energy | Recovery latency | Unrecoverable flip-flop errors |
|---|---|---|---|---|---|---|
| InO | Instruction Replay (IR) recovery | 16% | 21% | 21% | 47 cycles | None (all pipeline FFs recoverable) |
| | EIR recovery | 34% | 32% | 32% | 47 cycles | |
| | Flush recovery | 0.6% | 0.9% | 1.8% | 7 cycles | FFs after memory write stage |
| OoO | Instruction Replay (IR) recovery | 0.1% | 0.1% | 0.1% | 104 cycles | None (all pipeline FFs recoverable) |
| | EIR recovery | 0.2% | 0.1% | 0.1% | 104 cycles | |
| | Reorder Buffer (ROB) recovery | 0.01% | 0.01% | 0.01% | 64 cycles | FFs after reorder buffer stage |

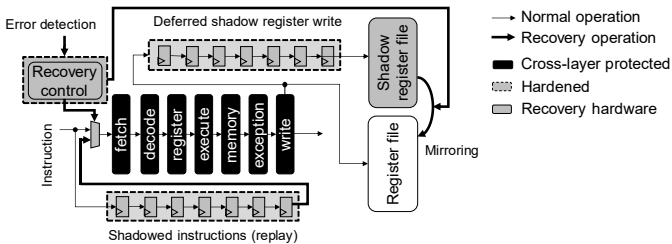

Figure 4. Instruction Replay (IR) recovery.

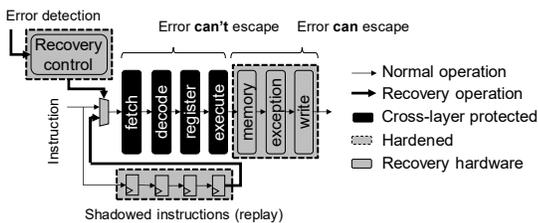

Figure 5. Flush recovery.

**Additional Techniques**: Many additional resilience techniques have been published in literature; but, these techniques are closely related to our evaluated techniques. Therefore, we believe that our results are representative and largely cover the cross-layer design space.

At the circuit-level, hardened flip-flops like DICE (Dual Interlocked storage Cell) [Calin 96], BCDMR (Bistable Cross-coupled Dual Modular Redundancy) [Furuta 10], and BISER (Built In Soft Error Resilience) [Mitra 05] are similar in cost to LEAP-DICE, the most resilient hardened flip-flop studied. The DICE technique suffers from an inability to tolerate SEMUs, unlike LEAP-DICE. BISER is capable of operating in both economy and resilient modes. This enhancement is provided by LEAP-ctrl. Hardened flip-flops like RCC (Reinforcing Charge Collection) [Seifert 10] offer around 5× soft error rate improvement at around 1.2× area, power, and energy cost. LHL provides slightly more soft error tolerance at roughly the same cost as RCC. Circuit-level detection techniques such as [Blaauw 08, Fojtik 13, Nicolaidis 99] are similar to EDS. Like EDS, these techniques can detect soft errors while offering minor differences in actual implementation. Stability checking [Franco 94] works on a similar principle of time sampling to detect errors.

Logic-level techniques like residue codes [Ando 03] can be effective for specific functional units like multipliers, but are costlier to implement than the simple XOR-trees used in logic parity. Additional logic level coding techniques like Berger codes [Berger 61] and Bose-Lin codes [Bose 85] are costlier to implement than logic parity. Like logic parity checking, residue, Berger, and Bose-Lin codes only detect errors.

Techniques like DMR (Dual Modular Redundancy) and TMR (Triple Modular Redundancy) at the architecture level can be easily ruled out since these techniques will incur more than 100% area, power, and energy costs. RMT (Redundant Multi-Threading) [Mukherjee 02] has been shown to have high (>40%) energy costs (which can increase due to recovery since RMT only serves to detect errors). Additionally, RMT is highly architecture dependent, which limits its applicability.

Software techniques like Shoestring [Feng 10], Error detectors [Pattabiraman 09], Reliability-driven transforms [Rehman 14], and SWIFT [Reis 05a] are similar to EDDI, but offer variations to the technique by reducing the number of checks added. As a result, EDDI can be used as a bound on the maximum error detection possible. An enhancement to SWIFT, known as CRAFT [Reis 05b], uses HW acceleration to improve reliability, but doing so eliminates the benefit of EDDI as a software-only technique. Although it is difficult to faithfully compare these "selective" EDDI techniques as published (since the original authors evaluated improvements using high-level error injection at the architecture register level which are generally inaccurate), the published results for these "selective" EDDI techniques show insufficient benefit (Table 16). Enhancements which reduce the execution time impact provide very low SDC improvements, while those that provide moderate improvement incur high execution time (and thus, energy) impact (much higher than providing the same improvement using LEAP-DICE, for instance). Fault screening [Racunas 07] is an additional software level technique. However, this technique also checks to ensure intermediate values computed during execution fall within expected bounds, which is similar to the mechanisms behind Software assertions for general-purpose processors, and thus, is covered by the latter.

Table 16. Comparison of "selective" EDDI techniques as reported in literature compared to EDDI evaluated using flip-flop-level error injection.

| | Error-injection | SDC improve | Exec. time impact |
|---|---|---|---|
| EDDI with store-readback (implemented) | Flip-flop | 37.8× | 2.1× |
| Reliability-aware transforms (published) | Arch. reg. | 1.8× | 1.05× |
| Shoestring (published) | Arch. reg. | 5.1× | 1.15× |
| SWIFT (published) | Arch. reg. | 13.7× | 1.41× |

**Low-level Techniques**: Resilience techniques at the circuit and logic layer (i.e., *low-level techniques*) are tunable as they can be selectively applied to individual flip-flops. As a result, a range of SDC/DUE improvements can be achieved for varying costs (Table 17). These techniques offer the ability to finely tune the specific flip-flops to protect in order to achieve the degree of resilience improvements required.

Table 17[15]. Costs vs. SDC and DUE improvements for tunable resilience techniques.
A (area cost %), P (power cost %), E (energy cost %)  (P=E for these combinations - no clock/execution time impact)

| | | | Bounded latency recovery | | | | | | | | | | Unconstrained recovery[16] | | | | | | | | | | Exec. time impact |
| | | | SDC improvement | | | | | DUE improvement | | | | | SDC improvement | | | | | DUE improvement | | | | | |
| | | | 2 | 5 | 50 | 500 | max | 2 | 5 | 50 | 500 | max | 2 | 5 | 50 | 500 | max | 2 | 5 | 50 | 500 | max | |
| InO | LEAP-DICE only | A<br>E | 0.8<br>2 | 1.8<br>4.3 | 2.9<br>7.3 | 3.3<br>8.2 | 9.3<br>22.4 | 0.7<br>1.5 | 1.7<br>3.8 | 3.8<br>9.5 | 5.1<br>12.5 | 9.3<br>22.4 | 0.8<br>2 | 1.8<br>4.3 | 2.9<br>7.3 | 3.3<br>8.2 | 9.3<br>22.4 | 0.7<br>1.5 | 1.7<br>3.8 | 3.8<br>9.5 | 5.1<br>12.5 | 9.3<br>22.4 | 0% |
| | Logic parity only (+ IR recovery) | A<br>E | 17.3<br>23.4 | 18.6<br>26 | 20.3<br>29.4 | 20.7<br>30.5 | 26.9<br>44.1 | 16.9<br>22.5 | 18.3<br>25.4 | 21.5<br>31.9 | 22.8<br>35 | 23.3<br>35.9 | 1.3<br>2.4 | 2.6<br>5 | 4.3<br>8.4 | 4.7<br>9.5 | 10.9<br>23.1 | - | - | - | - | - | 0% |
| | EDS-only (+ IR recovery) | A<br>E | 17.1<br>23.1 | 18.1<br>25.4 | 19.7<br>28.5 | 20.5<br>29.6 | 26.7<br>43.9 | 16.8<br>22.1 | 18<br>25.2 | 20.3<br>31.5 | 22.5<br>39.2 | 26.2<br>43.7 | 1.1<br>2.1 | 2.1<br>4.4 | 3.7<br>7.5 | 4.5<br>8.6 | 10.7<br>22.9 | - | - | - | - | - | 0% |
| OoO | LEAP-DICE only | A<br>E | 1.1<br>1.5 | 1.3<br>1.7 | 2.2<br>3.1 | 2.4<br>3.5 | 6.5<br>9.4 | 1.3<br>2 | 1.6<br>2.3 | 3.1<br>4.2 | 3.6<br>5.1 | 6.5<br>9.4 | 1.1<br>1.5 | 1.3<br>1.7 | 2.2<br>3.1 | 2.4<br>3.5 | 6.5<br>9.4 | 1.3<br>2 | 1.6<br>2.3 | 3.1<br>4.2 | 3.6<br>5.1 | 6.5<br>9.4 | 0% |
| | Logic parity only (+ IR recovery) | A<br>E | 1.9<br>1.6 | 2.1<br>2.4 | 6.1<br>4.1 | 6.3<br>5.1 | 14.2<br>13.7 | 1.7<br>2.4 | 2.6<br>3 | 4.5<br>4.4 | 5<br>5.4 | 13.8<br>13.6 | 1.8<br>1.5 | 2<br>2.3 | 5.9<br>4 | 6.2<br>5 | 14.1<br>13.6 | - | - | - | - | - | 0% |
| | EDS-only (+ IR recovery) | A<br>E | 1.4<br>1.7 | 1.8<br>2.1 | 3.3<br>3.5 | 4<br>4 | 12.3<br>11.6 | 1.3<br>2.1 | 2<br>2.5 | 3.6<br>4.4 | 4<br>5.3 | 11.8<br>11.4 | 1.3<br>1.6 | 1.7<br>2 | 3.2<br>3.4 | 3.9<br>3.9 | 12.2<br>11.5 | - | - | - | - | - | 0% |

**High-level Techniques**: In general, techniques at the architecture, software, and algorithm layers (i.e., *high-level techniques*) are less tunable as there is little control of the exact subset of flip-flops a high-level technique will protect. From Table 3, we see that no high level technique provides more than 38× improvement (while most offer far less improvement). As a result, to achieve a 50× improvement, for example, augmentation from low-level techniques at the circuit- and logic-level are required, regardless.

## 3. Cross-Layer Combinations

CLEAR uses a top-down approach to explore the cost-effectiveness of various cross-layer combinations. For example, resilience techniques at the upper layers of the system stack (e.g., ABFT correction) are applied before incrementally moving down the stack to apply techniques from lower layers (e.g., an optimized combination of logic parity checking, circuit-level LEAP-DICE, and micro-architectural recovery). This approach (example shown in Fig. 6) ensures that resilience techniques from various layers of the stack effectively interact with one another. Resilience techniques from the algorithm, software, and architecture layers of the stack generally protect multiple flip-flops (determined using error injection); however, a designer typically has little control over the specific subset protected. Using multiple resilience techniques from these layers can lead to situations where a given flip-flop may be protected (sometimes unnecessarily) by multiple techniques. At the logic and circuit layers, fine-grained protection is available since these techniques can be applied selectively to individual flip-flops (those not sufficiently protected by higher-level techniques).

We explore a total of 586 cross-layer combinations using CLEAR (Table 18). Not all combinations of the ten resilience techniques and four recovery techniques are valid (e.g., it is unnecessary to combine ABFT correction and ABFT detection since the techniques are mutually exclusive or to explore combinations of monitor cores to protect an InO-core due to the high cost). Accurate flip-flop level injection and layout evaluation reveals many individual techniques provide minimal (less than 1.5×) SDC/DUE improvement (contrary to conclusions reported in the literature that were derived using inaccurate architecture- or software-level injection), have high costs, or both. The consequence of this revelation is that most cross-layer combinations have high cost (detailed results for these costly combinations are omitted for brevity but are shown in Fig. 1).

Table 18. Creating 586 cross-layer combinations[2].

| | | No rec. | Flush / RoB rec. | IR / EIR rec. | Total |
|---|---|---|---|---|---|
| InO | Combinations of LEAP-DICE, EDS, parity, DFC, Assertions, CFCSS, EDDI | 127 | 3 | 14 | 144 |
| | ABFT correction / detection alone | 2 | 0 | 0 | 2 |
| | ABFT correction + previous combinations | 127 | 3 | 14 | 144 |
| | ABFT detection + previous combinations | 127 | 0 | 0 | 127 |
| | InO-core total | - | - | - | 417 |
| OoO | Combinations of LEAP-DICE, EDS, parity, DFC, monitor cores | 31 | 7 | 30 | 68 |
| | ABFT correction / detection alone | 2 | 0 | 0 | 2 |
| | ABFT correction + previous combinations | 31 | 7 | 30 | 68 |
| | ABFT detection + previous combinations | 31 | 0 | 0 | 31 |
| | OoO-core total | - | - | - | 169 |
| | Combined Total | | | | 586 |

### 3.1 Combinations for General-Purpose Processors

Among the 586 cross-layer combinations explored using CLEAR, a highly promising approach combines selective circuit-level hardening using LEAP-DICE, logic parity, and micro-architectural recovery (flush recovery for InO-cores, RoB recovery for OoO-cores). Thorough error injection using application benchmarks plays a critical role in selecting the flip-flops protected using these techniques. Figure 7 and Heuristic 1 detail the methodology for creating this combination. If recovery is not needed (e.g., for unconstrained recovery), the "Harden" procedure in Heuristic 1 can be modified to always return false.

For example, to achieve a 50× SDC improvement, the combination of LEAP-DICE, logic parity, and micro-architectural recovery provides a 1.5× and 1.2× energy savings for the OoO- and InO-cores, respectively, compared to selective circuit hardening using LEAP-DICE (Table 19). The relative benefits are consistent across benchmarks and over the range of SDC/DUE improvements. The overheads in Table 19 are small because we reported the most energy-efficient resilience solutions. Most of the 586 combinations are far costlier.

Let us consider the scenario where recovery hardware is not needed (e.g., unconstrained recovery). In this case, a minimal (<0.2% energy) savings can be achieved when targeting SDC improvement. However, without recovery hardware, DUEs increase since detected errors are now uncorrectable; thus, no DUE improvement is achievable.

Finally, one may suppose that the inclusion of EDS into cross-layer optimization may yield further savings since EDS costs ~25% less area,

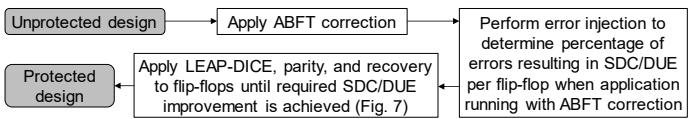

Figure 6. Cross-layer methodology example for combining ABFT correction, LEAP-DICE, logic parity, and micro-architectural recovery.

---

[15] Costs are generated per benchmark. We report the average cost over all benchmarks. Relative standard deviation is 0.6-3.1%.

[16] DUE improvements are not possible with detection-only techniques given unconstrained recovery.

Table 19. Costs vs. SDC and DUE improvements for various combinations in general-purpose processors.
A (area cost %), P (power cost %), E (energy cost %)

| | | | Bounded latency recovery | | | | | | | | | | Unconstrained recovery | | | | | | | | | | Exec. time impact |
|---|---|---|---|---|---|---|---|---|---|---|---|---|---|---|---|---|---|---|---|---|---|---|---|
| | | | SDC improvement | | | | | DUE improvement | | | | | SDC improvement | | | | | DUE improvement | | | | | |
| | | | 2 | 5 | 50 | 500 | max | 2 | 5 | 50 | 500 | max | 2 | 5 | 50 | 500 | max | 2 | 5 | 50 | 500 | max | |
| InO | LEAP-DICE + logic parity (+ flush recovery) | A<br>P<br>E | 0.7<br>1.9<br>1.9 | 1.7<br>3.9<br>3.9 | 2.5<br>6.1<br>6.1 | 3<br>6.7<br>6.7 | 8<br>17.9<br>17.9 | 0.6<br>1.5<br>1.5 | 1.5<br>3.4<br>3.4 | 3.6<br>8.4<br>8.4 | 4.4<br>10.4<br>10.4 | 8<br>17.9<br>17.9 | 0.7<br>1.9<br>1.9 | 1.6<br>3.8<br>3.8 | 2.4<br>5.9<br>5.9 | 2.8<br>6.5<br>6.5 | 7.6<br>17.2<br>17.2 | - | - | - | - | - | 0% |
| | EDS + LEAP-DICE + logic parity (+ flush recovery) | A<br>P<br>E | 0.9<br>1.9<br>1.9 | 2.3<br>4.3<br>4.3 | 2.7<br>6.6<br>6.6 | 3.3<br>7.2<br>7.2 | 8.4<br>19.3<br>19.3 | 0.8<br>1.7<br>1.7 | 2.1<br>3.8<br>3.8 | 3.8<br>8.5<br>8.5 | 4.8<br>11<br>11 | 8.4<br>19.3<br>19.3 | 0.9<br>1.9<br>1.9 | 2.2<br>4.2<br>4.2 | 2.5<br>6.3<br>6.3 | 3.2<br>7.1<br>7.1 | 8.1<br>19<br>19 | - | - | - | - | - | 0% |
| | DFC + LEAP-DICE + logic parity (+ EIR recovery) | A<br>P<br>E | 39.3<br>32.4<br>44.2 | 41.1<br>35.5<br>56.7 | 41.5<br>38.7<br>60.2 | 43.1<br>41<br>62.7 | 45<br>50.9<br>60.3 | 39.3<br>32.5<br>45.8 | 39.9<br>33.9<br>48.9 | 41.9<br>38.4<br>58.3 | 42.5<br>40.7<br>63 | 45<br>50.9<br>60.3 | 3.3<br>1.4<br>10.6 | 5.1<br>4.8<br>13.9 | 5.6<br>8.1<br>17.4 | 7.1<br>10<br>19.9 | 10.6<br>18.2<br>25.5 | - | - | - | - | - | 6.2% |
| | Assertions + LEAP-DICE + logic parity (no recovery) | A<br>P<br>E | - | - | - | - | - | - | - | - | - | - | 0.7<br>1.4<br>17.1 | 0.<br>1.8<br>17.5 | 1<br>2.2<br>18 | 1.1<br>2.2<br>18 | 7.6<br>17.2<br>24.5 | - | - | - | - | - | 15.6% |
| | CFCSS + LEAP-DICE + logic parity (no recovery) | A<br>P<br>E | - | - | - | - | - | - | - | - | - | - | 0.3<br>0.8<br>41.5 | 1<br>1.8<br>43 | 1.4<br>2.9<br>44.6 | 1.3<br>3.1<br>44.9 | 7.6<br>17.2<br>64.8 | - | - | - | - | - | 40.6% |
| | EDDI + LEAP-DICE + logic parity (no recovery) | A<br>P<br>E | - | - | - | - | - | - | - | - | - | - | 0<br>0<br>110 | 0<br>0<br>110 | 0.7<br>0.6<br>111 | 0.9<br>0.8<br>111 | 7.6<br>17.2<br>146 | - | - | - | - | - | 110% |
| OoO | LEAP-DICE + logic parity (+ ROB recovery) | A<br>P<br>E | 0.06<br>0.1<br>0.1 | 0.1<br>0.2<br>0.2 | 1.4<br>2.1<br>2.0 | 2.2<br>2.4<br>2.4 | 4.9<br>7<br>7 | 0.5<br>0.1<br>0.1 | 0.7<br>0.1<br>0.1 | 2.6<br>2<br>2 | 3<br>1.8<br>1.8 | 4.9<br>7<br>7 | 0.06<br>0.1<br>0.1 | 0.1<br>0.2<br>0.2 | 1.4<br>2.1<br>2.1 | 2.2<br>2.4<br>2.4 | 4.9<br>7<br>7 | - | - | - | - | - | 0% |
| | EDS + LEAP-DICE + logic parity (+ ROB recovery) | A<br>P<br>E | 0.07<br>0.1<br>0.1 | 0.1<br>0.2<br>0.2 | 1.6<br>2.3<br>2.3 | 2.2<br>2.5<br>2.5 | 5.4<br>8.1<br>8.1 | 0.6<br>0.1<br>0.1 | 0.8<br>0.1<br>0.1 | 2.6<br>2<br>2 | 3<br>1.8<br>1.8 | 5.4<br>8.1<br>8.1 | 0.07<br>0.1<br>0.1 | 0.1<br>0.2<br>0.2 | 1.6<br>2.3<br>2.3 | 2.2<br>2.5<br>2.5 | 5.4<br>8.1<br>8.1 | - | - | - | - | - | 0% |
| | DFC + LEAP-DICE + logic parity (+ EIR recovery) | A<br>P<br>E | 0.2<br>1.1<br>21.2 | 1<br>1.4<br>21.5 | 1.8<br>2<br>22.2 | 2<br>2.8<br>23 | 5.3<br>7.2<br>14.8 | 0.2<br>0.2<br>20 | 0.4<br>0.2<br>20.1 | 1.7<br>2.6<br>22.9 | 3.9<br>3.3<br>23.6 | 5.3<br>7.2<br>14.8 | 0.1<br>1<br>10 | 0.8<br>1.3<br>11.4 | 1.6<br>1.9<br>12.1 | 1.8<br>2.7<br>12.9 | 5.1<br>7.1<br>14.7 | - | - | - | - | - | 7.1% |
| | Monitor core + LEAP-DICE + logic parity (+ ROB rec.) | A<br>P<br>E | 9<br>16.3<br>16.3 | 9<br>16.3<br>16.3 | 9.8<br>20<br>20 | 10.5<br>20.2<br>20.2 | 13.9<br>23.3<br>23.3 | 9<br>16.3<br>16.3 | 9<br>16.3<br>16.3 | 10.1<br>20.1<br>20.1 | 11.2<br>21.5<br>21.5 | 13.9<br>22.3<br>22.3 | 9<br>16.3<br>16.3 | 9<br>16.3<br>16.3 | 9.8<br>20<br>20 | 10.5<br>20.2<br>20.2 | 13.9<br>23.3<br>23.3 | - | - | - | - | - | 0% |

power, energy than LEAP-DICE. However, a significant portion of EDS overhead is not captured solely by cell overhead. In fact, the additional cost of aggregating and routing the EDS error detection signals and the cost of adding delay buffers to satisfy minimum delay constraints posed by EDS dominates cost and prevents cross-layer combinations using EDS from yielding benefits (Table 19). Various additional cross-layer combinations spanning circuit, logic, architecture, and software layers are presented in Table 19.

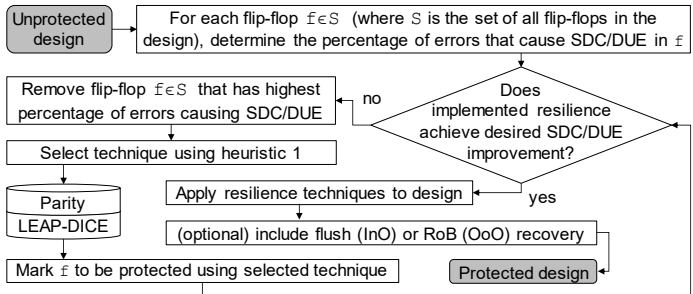

Figure 7. Cross-layer resilience methodology for combining LEAP-DICE, parity, and micro-architectural recovery.

---
**Heuristic 1:** Choose LEAP-DICE or parity technique
**Input:** `f`: flip-flop to be protected
**Output:** Technique to apply to `f` (LEAP-DICE, parity)
1:  **if** HARDEN(`f`) **then return** LEAP-DICE
2:  **if** PARITY(`f`) **then return** parity
3:  **return** LEAP-DICE

4:  **procedure** HARDEN(`f`)
5:    **if** an error in `f` cannot be flushed (i.e., `f` is in the memory, exception, writeback stages of InO or after the RoB of OoO)
6:    **then return** TRUE; **else return** FALSE
7:  **end procedure**

8:  **procedure** PARITY(`f`)
9:    **if** `f` has timing path slack greater than delay imposed by 32-bit XOR-tree (this implements low cost parity checking as explained in Sec. 2.4)
10:   **then return** TRUE, **else return** FALSE
11: **end procedure**

---

Up to this point, we have considered SDC and DUE improvements separately. However, it may be useful to achieve a specific improvement in SDC and DUE simultaneously. When targeting SDC improvement, DUE improvement also improves (and vice-versa); however, it is unlikely that the two improvements will be the same since flip-flops with high SDC vulnerability will not necessarily be the same flip-flops that have high DUE vulnerability. A simple method for targeting joint SDC/DUE improvement is to implement resilience until SDC (DUE) improvement is reached and then continue implementing resilience to unprotected flip-flops until DUE (SDC) improvement is also achieved. This ensures that both SDC and DUE improvement meet (or exceed) the targeted minimum required improvement. Table 20 details the costs required to achieve joint SDC/DUE improvement using this methodology when considering a combination of LEAP-DICE, parity, and flush/RoB recovery.

Table 20. Cost to achieve joint SDC/DUE improvement with a combination of LEAP-DICE, parity, and flush/RoB recovery.

| Joint SDC/DUE improvement | InO | | | OoO | | |
|---|---|---|---|---|---|---|
| | Area | Power | Energy | Area | Power | Energy |
| 2× | 0.7% | 2% | 2% | 0.6% | 0.1% | 0.1% |
| 5× | 1.9% | 4.2% | 4.2% | 0.9% | 0.4% | 0.4% |
| 50× | 4.1% | 9% | 9% | 2.8% | 2.2% | 2.2% |
| 500× | 4.6% | 10.8% | 10.8% | 3.1% | 2.8% | 2.8% |
| max | 8% | 17.9% | 17.9% | 4.9% | 7% | 7% |

### 3.2 Targeting Specific Applications

When the application space targets specific algorithms (e.g., matrix operations), a cross-layer combination of LEAP-DICE, parity, ABFT correction, and micro-architectural error recovery (flush/RoB) provides additional energy savings (compared to the general-purpose cross-layer combinations presented in Sec. 3.1. Since ABFT correction performs in-place error correction, no separate recovery mechanism is required for ABFT correction. For our study, we could apply ABFT correction to three of our PERFECT benchmarks: 2d_convolution, debayer_filter, and inner_product (the rest were protected using ABFT detection).

Table 21. Costs vs. SDC and DUE improvement for various cross-layer combinations involving ABFT.
A (area cost %), P (power cost %), E (energy cost %)

| | | | Bounded latency recovery | | | | | | | | | | Unconstrained recovery | | | | | | | | | | Exec. time impact |
|---|---|---|---|---|---|---|---|---|---|---|---|---|---|---|---|---|---|---|---|---|---|---|---|
| | | | SDC improvement | | | | | DUE improvement | | | | | SDC improvement | | | | | DUE improvement | | | | | |
| | | | 2 | 5 | 50 | 500 | max | 2 | 5 | 50 | 500 | max | 2 | 5 | 50 | 500 | max | 2 | 5 | 50 | 500 | max | |
| InO | ABFT correction + LEAP-DICE + logic parity (+ flush recovery) | A P E | 0 0 1.4 | 0.4 0.7 2.2 | 1.0 1.7 3.1 | 1.2 1.8 3.2 | 8 17.9 19.6 | 0.3 1 2.4 | 0.4 1 2.4 | 1.5 3.3 4.8 | 2.7 5.7 7.2 | 8 17.9 19.6 | 0 0 1.4 | 0.4 0.7 2.2 | 0.9 1.6 3 | 1.1 1.8 3.2 | 7.6 17.2 18.8 | - | - | - | - | - | 1.4% |
| | ABFT detection + LEAP-DICE + logic parity (no recovery) | A P E | - | - | - | - | - | - | - | - | - | - | 0 0 1.4 | 1.2 2.4 27 | 2 4.8 30 | 2.5 5.7 31.1 | 7.6 17.2 45.3 | - | - | - | - | - | 24% |
| | ABFT correction + LEAP-ctrl + LEAP-DICE + logic parity (+ flush recovery) | A P E | 1.5 0.6 1.9 | 2.5 1.3 2.7 | 3.8 2.6 4.0 | 4.1 2.8 4.2 | 8 17.9 19.6 | 1 1.3 2.8 | 1 1.3 2.8 | 4.1 4.6 6.1 | 5 7 8.5 | 8 17.9 19.6 | 1.5 0.6 1.9 | 2.3 1.2 2.6 | 3.4 2.6 4 | 4 2.7 4.1 | 7.6 17.2 18.8 | - | - | - | - | - | 1.4% |
| OoO | ABFT correction + LEAP-DICE + logic parity (+ ROB recovery) | A P E | 0 0 1.4 | 0.01 0.01 1.5 | 0.3 0.5 1.9 | 0.5 0.8 2.2 | 4.9 7 8.5 | 0.4 0.1 1.5 | 0.6 0.1 1.5 | 2.1 3 4.2 | 3 1.6 3 | 4.9 7 8.5 | 0 0 1.4 | 0.01 0.01 1.5 | 0.3 0.5 1.9 | 0.5 0.8 2.2 | 4.8 6.9 8.4 | - | - | - | - | - | 1.4% |
| | ABFT detection + LEAP-DICE + logic parity (no recovery) | A P E | - | - | - | - | - | - | - | - | - | - | 0 0 24 | 0.1 0.2 24.2 | 0.7 1.2 25.5 | 1.2 1.6 26 | 4.8 6.9 32.6 | - | - | - | - | - | 24% |
| | ABFT correction + LEAP-ctrl + LEAP-DICE + logic parity (+ ROB recovery) | A P E | 1.5 0.3 1.7 | 1.8 0.3 1.7 | 2.9 1.0 2.5 | 3.2 1.3 2.7 | 4.9 7 8.5 | 0.6 0.1 1.5 | 0.9 0.1 1.5 | 2.8 3 4.3 | 3.6 1.6 3.1 | 4.9 7 8.5 | 1.5 0.3 1.7 | 1.8 0.3 1.7 | 2.9 1 2.5 | 3.2 1.3 2.7 | 4.9 6.9 8.4 | - | - | - | - | - | 1.4% |

The results in Table 21 confirm that combinations of ABFT correction, LEAP-DICE, parity, and micro-architectural recovery provide up to 1.1× and 2× energy savings over the previously presented combination of LEAP-DICE, parity, and recovery when targeting SDC improvement for the OoO- and InO-cores, respectively. However, as will be discussed in Sec. 3.2.1, the practicality of ABFT is limited when considering general-purpose processors.

When targeting DUE improvement, including ABFT correction provides no energy savings for the OoO-core. This is because ABFT correction (along with most architecture and software techniques like DFC, CFCSS, and Assertions) performs checks at set locations in the program. For example, a DUE resulting from an invalid pointer access can cause an immediate program termination before a check is invoked. As a result, this DUE would not be detected by the resilience technique.

Although ABFT correction is useful for general-purpose processors limited to specific applications, the same cannot be said for ABFT detection (Table 21). Figure 8 shows that, since ABFT detection cannot perform in-place correction, ABFT detection benchmarks cannot provide DUE improvement (any detected error necessarily increases the number of DUEs). Additionally, given the lower average SDC improvement and generally higher execution time impact for ABFT detection algorithms, combinations with ABFT detection do not yield low-cost solutions.

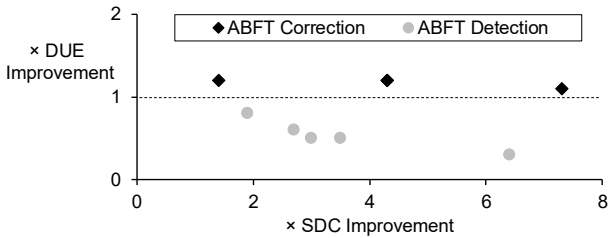

Figure 8. ABFT correction and ABFT detection benchmark comparison.

*3.2.1 Additional Considerations for ABFT*

Since most applications are not amenable to ABFT correction, the flip-flops protected by ABFT correction must also be protected by techniques such as LEAP-DICE or parity (or combinations thereof) for processors targeting general-purpose applications. This requires circuit hardening techniques (e.g., [Mitra 05, Zhang 06]) with the ability to selectively operate in an error-resilient mode (high resilience, high energy) when ABFT is unavailable, or in an economy mode (low resilience, low power mode) when ABFT is available. The LEAP-ctrl flip-flop accomplishes this task. The addition of LEAP-ctrl can incur an additional ~1% energy cost and ~3% area cost (Table 21).

Although 44% (22% for OoO-cores) of flip-flops would need to be implemented using LEAP-ctrl, only 5% (2% for OoO-cores) would be operating in economy mode at any given time (Table 22). Unfortunately, this requirement of fine-grained operating mode control is difficult to implement in practice since it would require some firmware or software control to determine and pass information to a hardware controller indicating whether or not an ABFT application were running and which flip-flops to place in resilient mode and which to place in economy mode (rather than a simple switch setting all such flip-flops into the same operating mode). Therefore, cross-layer combinations using ABFT correction may not be practical or useful in general-purpose processors targeting general applications.

Table 22. Impact of ABFT correction on flip-flops.

| Core | % FFs with an error corrected by any ABFT algorithm (∪) | % FFs with an error corrected by every ABFT algorithm (∩) |
|---|---|---|
| InO | 44% | 5% |
| OoO | 22% | 2% |

## 4. Application Benchmark Dependence

The most cost-effective resilience techniques rely on selective circuit hardening / parity checking guided by error injection using application benchmarks. This raises the question: what happens when the applications in the field do not match application benchmarks? We refer to this situation as *application benchmark dependence*.

To quantify this dependence, we randomly selected 4 (of 11) SPEC benchmarks as a *training set*, and used the remaining 7 as a *validation set*. Resilience is implemented using the training set and the resulting design's resilience is determined using the validation set. Therefore, the training set tells us which flip-flops to protect and the validation set allows us to determine what the actual improvement would be when this same set of flip-flops is protected. We used 50 training/validation pairs.

Since high-level techniques cannot be tuned to achieve a given resilience improvement, we analyze each as a standalone technique to better understand how they perform individually. For standalone resilience techniques, the average inaccuracy between the results of trained and validated resilience is generally very low (Table 23 and Table 24) and is likely due to the fact that the improvements that the techniques themselves provide is already very low. We also report p-values [Wasserstein 16], which provide a measure of how likely the validated improvement and trained improvement would match.

Table 23. Trained vs. validated SDC improvement for high-level techniques. Underestimation low because improvements are already low.

| Core | Technique | Train | Validate | Underestimate | p-value |
|---|---|---|---|---|---|
| InO | DFC | 1.3× | 1.2× | -7.7% | 3.8×10$^{-9}$ |
| | Assertions | 1.5× | 1.4× | -6.7% | 2.4×10$^{-1}$ |
| | CFCSS | 1.6× | 1.5× | -6.3% | 5.7×10$^{-1}$ |
| | EDDI | 37.8× | 30.4× | -19.6% | 6.9×10$^{-1}$ |
| | ABFT correction | 4.3× | 3.9× | -9.3% | 6.7×10$^{-1}$ |
| OoO | DFC | 1.3× | 1.2× | -7.7% | 1.9×10$^{-5}$ |
| | Monitor core | 19.6× | 17.5× | -5.6% | 8.3×10$^{-3}$ |
| | ABFT correction | 4.3× | 3.7× | -14% | 7.2×10$^{-1}$ |

Table 24. Trained vs. validated DUE improvement for high-level techniques. Underestimation low because improvements are already low.

| Core | Technique | Train | Validate | Underestimate | p-value |
|---|---|---|---|---|---|
| InO | DFC | 1.4× | 1.3× | -7.1% | 3.9×10$^{-17}$ |
| | Assertions | 0.6× | 0.6× | 0% | 8×10$^{-2}$ |
| | CFCSS | 0.6× | 0.6× | 0% | 9.2×10$^{-1}$ |
| | EDDI | 0.4× | 0.4× | 0% | 2.2×10$^{-1}$ |
| | ABFT correction | 1.2× | 1.2× | 0% | 1.8×10$^{-1}$ |
| OoO | DFC | 1.4× | 1.3× | -7.1% | 1.4×10$^{-10}$ |
| | Monitor core | 15.2× | 13.9× | -8.6% | 3.5×10$^{-7}$ |
| | ABFT correction | 1.1× | 1.1× | 0% | 1.5×10$^{-1}$ |

Table 25 and Table 26 indicate that validated SDC and DUE improvements are generally underestimated. Fortunately, when targeting <10× SDC improvement, the underestimation is minimal. This is due to the fact that the most *vulnerable* 10% of flip-flops (i.e., the flip-flops that result in the most SDCs or DUEs) are consistent across benchmarks. Since the number of errors resulting in SDC or DUE is not uniformly distributed among flip-flops, protecting these top 10% of flip-flops will result in the ~10× SDC improvement regardless of the benchmark considered. The vulnerabilities of the remaining 90% of flip-flops are more benchmark-dependent. Concretely, we can analyze benchmark similarity by analyzing the vulnerable flip-flops indicated by each application benchmark. Per benchmark, one can group the most vulnerable 10% of flip-flops into a subset (e.g., subset 1). The next 10% of vulnerable flip-flops (e.g., 10-20%) are grouped into subset 2 (and so on up to subset 10). Therefore, given our 18 benchmarks, we create 18 distinct subset 1's, 18 distinct subset 2's, and so on. Each group of 18 subsets (e.g., all subset 1's) can then be assigned a similarity as given in Eq. 2. The similarity of subset "x" is the number of flip-flops that exist in all subset "x"s (e.g., subset intersection) divided by the number of unique flip-flops in every subset "x"s (e.g., subset union). From Table 27, it is clear that only the top 10% most vulnerable flip-flops have very high commonality across all benchmarks (the last 2 subsets have high similarity because these are the flip-flops that have errors that always vanish). All other flip-flops are relatively distributed across the spectrum depending on the specific benchmark being run.

$$Simularity\ (subset\ "x") = \frac{|\cap (all\ flip-flops\ in\ every\ subset\ "x")|}{|\cup (all\ flip-flops\ in\ every\ subset\ "x")|} \quad (Eq.\ 2)$$

It is clear that for highly-resilient designs, one must develop methods to combat this sensitivity to benchmarks. Benchmark sensitivity may be minimized by training using additional benchmarks or through better benchmarks (e.g., [Mirkhani 15a]). An alternative approach is to apply our CLEAR framework using available benchmarks, and then replace all remaining unprotected flip-flops using LHL (Table 4). This enables our resilient designs to meet (or exceed) resilience targets at ~1% additional cost for SDC and DUE improvements >10×.

Table 25. SDC improvement, cost before and after applying LHL to otherwise unprotected flip-flops.

| Core | SDC improvement | | | Cost before LHL insertion | | Cost after LHL insertion | |
|---|---|---|---|---|---|---|---|
| | Train | Validate | After LHL | Area | Power / Energy | Area | Power / Energy |
| InO | 5× | 4.8× | 19.3× | 1.6% | 3.6% | 3.1% | 5.7% |
| | 10× | 9.6× | 38.2× | 1.7% | 3.9% | 3.1% | 5.7% |
| | 20× | 19.1× | 75.8× | 1.9% | 4.4% | 3.2% | 6.1% |
| | 30× | 26.8× | 105.6× | 2.2% | 4.8% | 3.2% | 6.3% |
| | 40× | 32.9× | 129.4× | 2.3% | 5.3% | 3.3% | 6.7% |
| | 50× | 38.9× | 152.3× | 2.4% | 5.7% | 3.3% | 6.9% |
| | 500× | 433.1× | 1,326.1× | 2.9% | 6.3% | 3.4% | 7.1% |
| | Max | 5,568.9× | 5,568.9× | 8% | 17.9% | 8% | 17.9% |
| OoO | 5× | 4.8× | 35.1× | 0.1% | 0.2% | 0.9% | 1.8% |
| | 10× | 8.8× | 40.7× | 0.4% | 0.6% | 1.1% | 2.1% |
| | 20× | 18.8× | 65.6× | 0.7% | 1% | 1.3% | 2.3% |
| | 30× | 21.3× | 82.3× | 0.9% | 1.4% | 1.4% | 2.4% |
| | 40× | 26.4× | 130.2× | 1.2% | 1.7% | 1.7% | 2.5% |
| | 50× | 32.1× | 204.3× | 1.4% | 2.1% | 1.9% | 2.7% |
| | 500× | 301.4× | 1084.1× | 2.2% | 2.4% | 2.4% | 2.8% |
| | Max | 6,625.8× | 6,625.8× | 4.9% | 7% | 4.9% | 7% |

Table 26. DUE improvement, cost before and after applying LHL to otherwise unprotected flip-flops.

| Core | DUE improvement | | | Cost before LHL insertion | | Cost after LHL insertion | |
|---|---|---|---|---|---|---|---|
| | Train | Validate | After LHL | Area | Power / Energy | Area | Power / Energy |
| InO | 5× | 4.7× | 18.7× | 1.5% | 3.4% | 3.3% | 5.9% |
| | 10× | 8.7× | 34.6× | 1.9% | 4.2% | 3.5% | 6.5% |
| | 20× | 16.3× | 64.5× | 2.4% | 5.3% | 3.7% | 7% |
| | 30× | 23.5× | 92.7× | 2.8% | 6.6% | 3.7% | 8.1% |
| | 40× | 29.9× | 117.6× | 3.3% | 7.5% | 4.1% | 8.7% |
| | 50× | 35.9× | 140.6× | 3.6% | 8.4% | 4.2% | 9.4% |
| | 500× | 243.5× | 840.3× | 4.4% | 10.4% | 4.8% | 10.9% |
| | Max | 5,524.7× | 5,524.7× | 8% | 17.9% | 8% | 17.9% |
| OoO | 5× | 4.4× | 28.7× | 0.7% | 0.1% | 1.8% | 1.7% |
| | 10× | 8.7× | 36.6× | 1.1% | 0.5% | 2.1% | 2% |
| | 20× | 17.3× | 70.2× | 1.5% | 0.9% | 2.5% | 2% |
| | 30× | 22.2× | 81.5× | 1.8% | 1.3% | 2.6% | 2.1% |
| | 40× | 26.1× | 115.1× | 2.1% | 1.6% | 2.8% | 2.4% |
| | 50× | 29.8× | 121.3× | 2.5% | 2% | 3.1% | 2.6% |
| | 500× | 153.2× | 625.1× | 2.9% | 1.9% | 3.4% | 2.7% |
| | Max | 6,802.6× | 6,802.6× | 4.9% | 7% | 4.9% | 7% |

Table 27. Subset similarity across all 18 benchmarks for the InO-core (subsets broken into groups consisting of 10% of all flip-flops).

| Subset (ranked by decreasing SDC + DUE vulnerability) | Similarity (Eq. 2) |
|---|---|
| 1: 0-10% | 0.83 |
| 2: 10-20% | 0.05 |
| 3: 20-30% | 0 |
| 4: 30-40% | 0 |
| 5: 40-50% | 0 |
| 6: 50-60% | 0 |
| 7: 60-70% | 0 |
| 8: 70-80% | 0 |
| 9: 80-90% | 0.71 |
| 10: 90-100% | 1 |

## 5. The Design of New Resilience Techniques

CLEAR has been used to comprehensively analyze the design space of existing resilience techniques (and their combinations). As new resilience techniques are proposed, CLEAR can incorporate and analyze these techniques as well. However, CLEAR can also be used today to guide the design of new resilience techniques.

All resilience techniques will lie on a two-dimensional plane of energy cost vs. SDC improvement (Fig. 9). The range of designs formed using combinations of LEAP-DICE, parity, and micro-architectural recovery form the lowest-cost cross-layer combination available using today's resilience techniques. In order for new resilience techniques to be able to create competitive cross-layer combinations, they must have energy and improvement tradeoffs that place the technique under the region bounded by our LEAP-DICE, parity, and micro-architectural recovery solution. Since certain standalone techniques, like LEAP-DICE, can also provide highly competitive solutions, it is useful to understand the cost vs. improvement tradeoffs for new techniques in relation to this best standalone technique as well (Fig. 10).

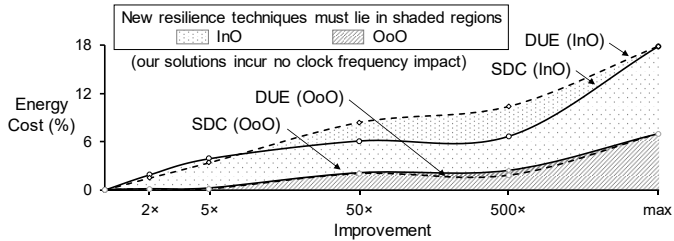

Figure 9. New resilience techniques must have cost and improvement tradeoffs that lie within the shaded regions bounded by LEAP-DICE + parity + micro-architectural recovery.

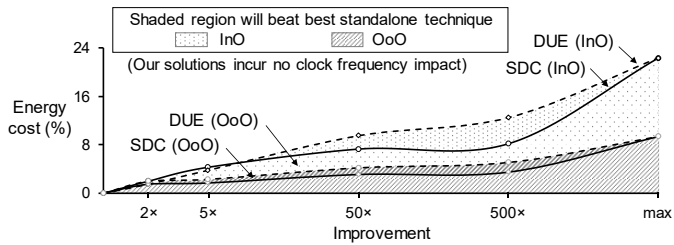

Figure 10. To be competitive standalone solutions, new resilience techniques must have cost and improvement tradeoffs that lie within the shaded regions bounded by LEAP-DICE.

## 6. Conclusions

CLEAR is a first of its kind cross-layer resilience framework that enables effective exploration of a wide variety of resilience techniques and their combinations across several layers of the system stack. Extensive cross-layer resilience studies using CLEAR demonstrate:

1. A carefully optimized combination of selective circuit-level hardening, logic-level parity checking, and micro-architectural recovery provides a highly cost-effective soft error resilience solution for general-purpose processors.

2. Selective circuit-level hardening alone, guided by thorough analysis of the effects of soft errors on application benchmarks, also provides a cost-effective soft error resilience solution (with ~1% additional energy cost for a 50× SDC improvement compared to the above approach).

3. Algorithm Based Fault Tolerance (ABFT) correction combined with selective circuit-level hardening (and logic-level parity checking and micro-architectural recovery) can further improve soft error resilience costs. However, existing ABFT correction techniques can only be used for a few applications; this limits the applicability of this approach in the context of general-purpose processors.

4. Based on our analysis, we can derive bounds on energy costs vs. degree of resilience (SDC or DUE improvements) that new soft error resilience techniques must achieve to be competitive.

5. It is crucial that the benefits and costs of new resilience techniques are evaluated thoroughly and correctly before publication. Detailed analysis (e.g., flip-flop-level error injection or layout-level cost quantification) identifies hidden weaknesses that are often overlooked.

While this paper focuses on soft errors in processor cores, cross-layer resilience solutions for accelerators and uncore components as well as other error sources (e.g., voltage noise) may have different tradeoffs and may require additional modeling and analysis capabilities.

## 7. Acknowledgment

This research is supported in part by DARPA MTO (contract no. HR0011-13-C-0022), DTRA, NSF, and SRC. The views expressed are those of the authors and do not reflect the official policy or position of the Department of Defense or the U.S. Government. Distribution Statement "A" (Approved for Public Release, Distribution Unlimited). We thank Prof. S. V. Adve (UIUC), Prof. M. Alioto (NUS), Dr. K. A. Bowman (Qualcomm), Dr. V. Chandra (ARM), Dr. B. Cline (ARM), Dr. S. Das (ARM), Dr. S. K. S. Hari (NVIDIA), Prof. J. C. Hoe (CMU), Prof. A. B. Kahng (UCSD), Dr. J. Kao (TSMC), Dr. D. Lin (Stanford), Prof. I. L. Markov (U. Michigan), Dr. H. Naeimi (Intel), Dr. A. S. Oates (TSMC), Prof. K. Pattabiraman (UBC), and Dr. S. K. Sahoo (UIUC). We thank the Texas Advanced Computing Center at The University of Texas at Austin.

## Appendix A. Flip-flops with Errors That Always Vanish

Flip-flops in 50 structures of the InO-core and 142 structures of the OoO-core have errors that always vanish.

**InO-core:** `a.ctrl.inst, a.ctrl.tt, a.ctrl.wy, a.cwp, a.rfe1, a.rfe2, d.pv, e.ctrl.inst, e.ctrl.tt, e.ctrl.wy, e.cwp, e.et, e.mac, e.mul, e.mulstep, e.su, e.ymsb, m.ctrl.inst, m.ctrl.pc, m.ctrl.tt, m.ctrl.wicc, m.ctrl.wy, m.dci.asi, m.dci.lock, m.dci.signed, m.irqen, m.irqen2, m.y, w.s.dwt, w.s.ec, w.s.ef, w.s.icc, w.s.pil, w.s.ps, w.s.tba, w.s.tt, w.s.y, x.ctrl.inst, x.ctrl.pc, x.ctrl.pv, x.ctrl.rett, x.ctrl.tt, x.ctrl.wicc, x.ctrl.wy, x.debug, x.icc, x.intack, x.ipend, x.npc, x.y`

**OoO-core:** `D0R0.reg0, D0R0.reg1, D0R0.reg2, D0R0.reg3, exec.ca0.br, exec.ca0.p0, exec.ca0.p1, exec.ca0.p2, exec.cb0.buffer.valid.reg, exec.cb0.queue.head.reg, exec.cb0.queue.tail.reg, exec.mu0.a01, exec.mu0.a12, exec.mu0.a23, exec.mu0.a34, exec.mu0.b01, exec.mu0.b12, exec.mu0.b23, exec.mu0.b34, exec.mu0.i0, exec.mu0.i1, exec.mu0.i2, exec.mu0.i3, F0.flushPC.reg, F1.reg0, F1.reg2, F1.reg3, F1.reg6, mem.finished.st2.reg, mem.l1dcache.accessaddr0.reg, mem.l1dcache.accessaddr1.reg, mem.l1dcache.accessaddrtype0.reg, mem.l1dcache.accessaddrtype1.reg, mem.l1dcache.accessfulldata0.reg, mem.l1dcache.accessfulldata1.reg, mem.l1dcache.accesshit0.reg, mem.l1dcache.addr.in0.reg, mem.l1dcache.addr.in1.reg, mem.l1dcache.addr.in2.reg, mem.l1dcache.addr.in3.reg, mem.l1dcache.addr.in5.reg, mem.l1dcache.addr.in7.reg, mem.l1dcache.addr1.out.reg, mem.l1dcache.addr2.out.reg, mem.l1dcache.data.in0.reg, mem.l1dcache.data.in1.reg, mem.l1dcache.data.in2.reg, mem.l1dcache.data.in3.reg, mem.l1dcache.data.in4.reg, mem.l1dcache.data.in5.reg, mem.l1dcache.data.in6.reg, mem.l1dcache.data.in7.reg, mem.l1dcache.data2.out.reg, mem.l1dcache.missqueue.delayed.returnedaddr1.reg, mem.l1dcache.missqueue.delayed.returnedaddr2.reg, mem.l1dcache.missqueue.delayed.returnedhit1.reg, mem.l1dcache.missqueue.delayed.returnedhit2.reg, mem.l1dcache.missqueue.q.done.f.reg, mem.l1dcache.missqueue.q.type.f.reg, mem.l1dcache.mobid2.out.reg, mem.l1dcache.size1.out.reg, mem.l1dcache.size2.out.reg, mem.l1dcache.write.in0.reg, mem.l1dcache.write.in1.reg, mem.l1dcache.write.in2.reg, mem.l1dcache.write.in3.reg, mem.l1dcache.write.in4.reg, mem.l1dcache.write.in5.reg, mem.l1dcache.write.in6.reg, mem.l1dcache.write.in7.reg, mem.ldq.address.phys.f.reg, mem.ldq.forward.f.reg, mem.ldq.num.entries.f.reg, mem.returned.hintvalid1.delayed.reg, mem.stb.forward.data1.delayed.reg, mem.stb.forward.data2.delayed.reg, mem.stb.forward.stid1.delayed.reg, mem.stb.forward.stid2.delayed.reg, mem.stq.address.v.f.reg, regs.ex.wb.agen1, regs.ex.wb.cond0, regs.ex.wb.d0, regs.ex.wb.data, regs.ex.wb.i0, regs.ex.wb.i1, regs.ex.wb.i2, regs.ex.wb.i3, regs.ex.wb.i4, regs.ex.wb.i5, regs.rr.ex.cond, regs.rr.ex.data, regs.rr.ex.i0, regs.rr.ex.i1, regs.rr.ex.i2, regs.rr.ex.i3, regs.rr.ex.i4, regs.rr.ex.i5, regs.rr.ex.valid, regs.wb.wb.brdir, regs.wb.wb.flushpc, regs.wb.wb.ret1, regs.wb.wb.ret2, regs.wb.wb.ret3, regs.wb.wb.ret4, regs.wb.wb.ret5, regs.wb.wb.ret6, regs.wb.wb.ret7, regs.wb.wb.ret8, regsIRR.reg0, regsIRR.reg1, regsIRR.reg2, regsIRR.reg3, regsIRR.reg4, regsIRR.reg5, regsR0R1.reg0, regsR0R1.reg1, regsR0R1.reg2, regsR0R1.reg3, regsR1I.reg0, regsR1I.reg1, regsR1I.reg2, regsR1I.reg3, RF0.F1.lhist, RF0.F1.ras.ret.inv, RF0.F1.takenAddress, RF0.PCreg, RF1.F2.inst0, RF1.F2.inst1, RF1.F2.inst2, RF1.F2.inst3, RF1.F2.inst4, RF1.F2.inst5, RF1.F2.inst6, RF1.F2.inst7, RF2.D0.reg0, RF2.D0.reg1, RF2.D0.reg2, RF2.D0.reg3, rob.rob.dir.f.reg, sched0.inst.array.reg, sched0.sb0.sb.activate.reg, sched0.sb0.sb.counters.reg`